\documentclass[prl,aps, twocolumn]{revtex4-1}
\usepackage{graphicx}
\usepackage{color}
\usepackage[tbtags]{amsmath}
\usepackage{amssymb}
\usepackage{bm}

\newcommand{\be}{\begin{equation}}
\newcommand{\ee}{\end{equation}}
\newcommand{\ba}{\[\begin{aligned}}
\newcommand{\ea}{\end{aligned}\]}
\newcommand{\HH}{{\cal H}}

\newcommand{\p}{\partial}
\newcommand{\s}{\sigma}
\newcommand{\la}{\langle}
\newcommand{\ra}{\rangle}

\newcommand{\lp}{\left(}
\newcommand{\rp}{\right)}

\renewcommand{\Im}{{\textrm{ Im}\,}}
\renewcommand{\vec}[1]{{\bm #1}}

\renewcommand{\phi}{\varphi}
\renewcommand{\epsilon}{\varepsilon}

\renewcommand{\dag}{\dagger}

\begin{document} 

\title{Soft pair excitations and double-log divergences due to carrier interactions in graphene} 

\author{Cyprian Lewandowski, L. S. Levitov}
\address{Department of Physics, Massachusetts Institute of Technology, 77 Massachusetts Avenue, Cambridge, MA02139}

\begin{abstract}
Interactions between charge carriers in graphene lead to logarithmic renormalization of observables mimicking the behavior known in (3+1)-dimensional quantum electrodynamics (QED). Here we analyze soft electron-hole (e-h) excitations generated as a result of fast charge dynamics, a direct analog of the signature QED effect---multiple soft photons produced by the QED vacuum shakeup. We show that such excitations are generated in photon absorption, when a photogenerated high-energy e-h pair cascades down in energy and gives rise to multiple soft e-h excitations. This fundamental process is manifested in a double-log divergence in the emission rate of soft pairs and a characteristic power-law divergence in their energy spectrum 
of the form $\frac{1}{\omega}\ln \lp\frac{\omega}{\Delta}\rp $. Strong carrier-carrier interactions make pair production a prominent pathway in the photoexcitation cascade. 
\end{abstract}

\maketitle

Low-energy electronic excitations in graphene combine, in a unique way, some aspects of two-dimensional (2D) and three-dimensional (3D) systems \cite{gonzalez1994,gonzalez1999,vafek2007,son2007}.  Namely,  charge carriers in  graphene sheets are described by a (2+1)-dimensional massless Dirac Hamiltonian, %reside in a two-dimensional (2D) graphene sheet 
while interactions between carriers are governed by a 3D Coulomb $1/r$ potential. The latter arises due to electric field that extends in the 3D space around graphene and remains unscreened at large distances. As a result, the long-wavelength theory of carrier interactions, sometimes called graphene quantum electrodynamics (GQED), has little in common with (2+1)-dimensional QED; instead it strongly resembles (3+1)-dimensional QED.  Similar to the latter, the diagrammatic expansion carried out in powers of dimensionless coupling is beset by log divergences. Furthermore, the renormalization group flow is towards a weak-coupling fixed point at long wavelengths, similar to (3+1)-dimensional QED. 

The renormalization scheme employed in GQED comes in two distinct flavors \cite{gonzalez1994,gonzalez1999,vafek2007,son2007,drut2008,foster2008,dassarma2013}. One is the weak-coupling approach in which a fine-structure constant 
%$\alpha = e^2/\hbar v\kappa$, with $\kappa$ a dielectric constant, 
is used as a perturbation parameter. The other is a ``strong-coupling'' diagrammatic expansion carried out in powers of the dynamically screened (RPA) interaction, corresponding to the expansion parameter $1/N$, where $N=4$ is the number of spin/valley flavors. The strong-coupling approach is usually taken to be more accurate than the weak-coupling approach \cite{hofmann2014} since the %fine structure %constant 
dimensionless interaction strength $\alpha = e^2/\hbar v\kappa$, with $\kappa$ the dielectric constant, %and %Dirac Fermi 
%the carrier velocity $300$ times smaller than the speed of light,  $v=c/300$, takes values of order 
typically exceeds unity. Various manifestations of log divergences %and renormalization in graphene QED 
in GQED have been studied, in particular renormalization of carrier velocity \cite{gonzalez1994,gonzalez1999,son2007,vafek2007, elias2011,dassarma2013}, Dirac mass \cite{kotov2009,song2013}, and the vertices describing coupling to external fields \cite{mishchenko2007,mishchenko2008,sheehy2009,juricic2010}. 
%LL \addLL{[MORE REFS]} 
These results, along with the recent studies of higher-order contributions to diagrammatic expansion \cite{barnes2014}, %indicate that 
helped to build a compelling case for %the strong coupling 
renormalization approach in GQED.

\begin{figure}
\includegraphics[width=0.99\linewidth]{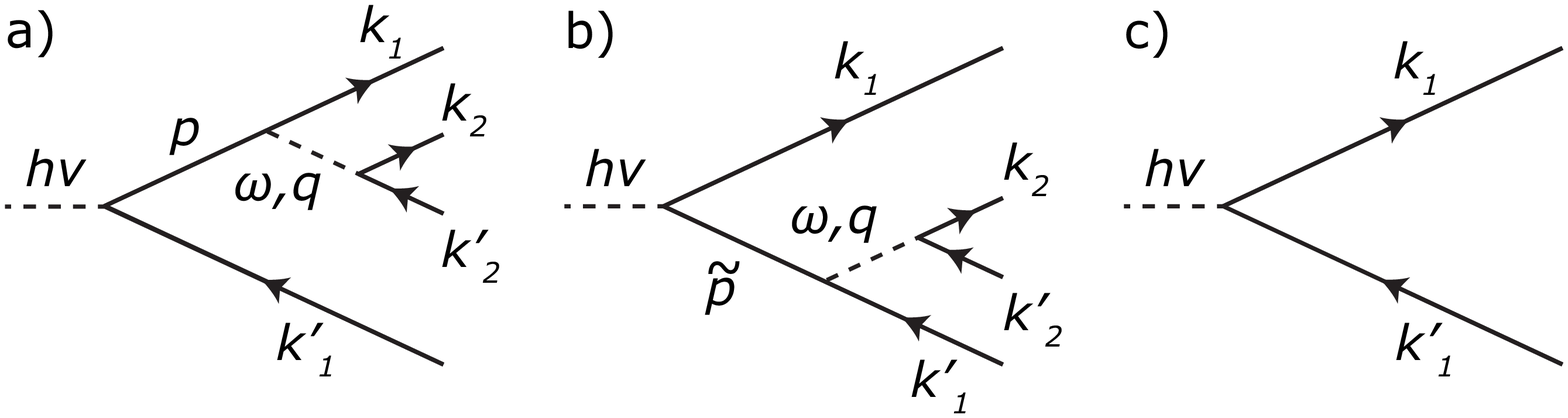} %{fig1abc} %{cones.eps}
%:
\caption{(a), (b) Diagrams describing soft pair creation following a hard pair photoexcitation. (c) Diagram describing photon absorption with no secondary pairs created. Lines with arrows denote fermion propagators; the outward and inward arrows correspond to creation of particles in the conduction band and holes in the valence band. The carrier-carrier interaction (inner dashed lines) is of the RPA form, Eq. \eqref{eq_RPA_V_Pi}.} 
\label{fig1} 
\end{figure}

Here we will analyze another interesting % somewhat more esoteric 
QED-related effect: creation of soft excitations %electron-hole pairs 
as a result of fast charge dynamics. This process represents a direct analog of a signature QED effect---Bremsstrahlung radiation due to fast particle dynamics in which multiple soft photons are emitted. %We recall that the 
The cross section describing soft photon emission is related to the %inclusive 
bare cross section through the seminal Sudakov double-log dependence \cite{peskin_schroeder},
\be\label{eq:sudakov}
d\sigma(p\to p'+{\rm photon}) %\hbar\omega)
=d\sigma(p\to p')\frac{\alpha}{\pi}\ln\frac{-q^2}{\mu^2}\ln\frac{-q^2}{m^2},
\ee
where $-q^2\to \infty$ is the fast particle momentum change, $m$ is electron mass and $\mu$ is the ``photon mass'', a parameter introduced to control the infrared divergence due to emission of soft photons. 
%LL \addLL{Generalization to QCD and the structure of \addLL{jets} in HEP.} 
%As discussed elsewhere\cite{lewandowski2017}, 
An analog of this process in graphene QED, discussed below,  is the emission of soft electron-hole (e-h) pairs accompanying fast charge dynamics. In particular, processes of this type are expected to occur during photoexcitation when a high-energy e-h pair, created through photoabsorption, cascades down in energy giving rise to multiple soft e-h excitations. The corresponding diagrams are pictured in Fig. \ref{fig1}.

Apart from a difference in the nature of excitations, soft e-h pairs vs soft photons, the process 
%energy spectrum 
of soft pair production in GQED %as a result of  photo-absorption in graphene 
bears a strong similarity to Eq. \eqref{eq:sudakov}. 
Both %dependences 
arise at first order in the dimensionless coupling and feature a double-log dependence. 
Since carrier dispersion in graphene is gapless, both logs are IR divergent and require regularization by implementing some form of cutoff at low-energy scales.  
We will start by considering a cutoff due to a small mass gap of Dirac carriers through replacing the linear dispersion $\epsilon=\pm v|\vec k|$ with $\epsilon=\pm \sqrt{v^2\vec k^2+\Delta^2}$. 
Accounting for the cutoff %at the energy $\Delta$ 
we find 
a power-law spectrum of emitted soft pairs: %soft pairs produced as a result of  photo-absorption to be of 
%form
%
\be\label{eq:spectrum}
dW(\omega)=W_0 \frac{16 }{\pi^2 N}  \, \frac{\ln\frac{\omega}{\Delta}}{\omega} d\omega
,\quad
\Delta<\omega<h\nu/2
,
\ee
where $h\nu$ is  photon energy, $W_0$ is the photoabsorption rate without soft pair creation, and the log divergence is regulated by $\Delta$. As in QED, the single pair emission rate in Eq. \eqref{eq:spectrum} can be reinterpreted in terms of the expected number of emitted pairs provided that different pairs are statistically independent \cite{peskin_schroeder}. % and thus have a Poisson distribution. 
Integration over $\Delta<\omega< h\nu/2$ gives the mean number of photogenerated pairs, 
\be\label{eq:Nave}
\la N_{\rm p}\ra=\frac1{W_0}\int_\Delta^{h\nu/2}d W(\omega) %d\omega
=
\frac{8}{\pi^2 N} \lp \ln\frac{h\nu}{2\Delta} \rp^2, %, \log\frac{h\nu}{2\Delta_H}.
\ee 
which is much greater than unity at small $\Delta\ll  h\nu/2$ and, as evident from Eq. \eqref{eq:spectrum}, is dominated by soft excitations with low energies $\omega\ll h\nu$. 

We will also consider another cutoff mechanism, arising due to screening of the long-range part of  $1/r$ interaction by a gate placed a distance $H$ away from the graphene plane. Screening suppresses excitations with wavenumbers smaller that $k_H=1/(2H)$, % where $H$ is the distance to the gate. 
%These wavenumbers 
which translates into a cutoff energy scale $\Delta_H=\hbar v k_H$. 
Interestingly, while this cutoff impacts both log functions in the double-log dependence, only one of them becomes finite whereas the other one must be regulated by a second cutoff mechanism such as the Dirac gap $\Delta$ in Eqs. \eqref{eq:spectrum} and \eqref{eq:Nave}. The resulting double-log dependence, Eq. \eqref{eq:Np=loglog}, has two different cutoffs: one at $\Delta_H$, another at $\mu=\Delta^2/\Delta_H$. % set by $\Delta_H$ and $\mu=\Delta^2/\Delta_H$, respectively. 
This behavior, as well as the fact that the double-log dependence appears at first order in dimensionless coupling $1/N$, %are in full analogy with 
accurately mimics the properties of the soft photon emission in QED. 

%LL \addLL{Relevance: sources of hot electrons; cite Qiong's paper; carrier multiplication}

\section{Model of photoabsorption}

Electrons in graphene are described by the Hamiltonian for $N$ species of massless Dirac particles in a plane coupled by a long-range $1/r$ interaction due to electric field in 3D space:
\be\label{eq:H=H0+V}
\HH\! =\! \int\! d^2x \sum_{i=1}^N \psi^{\dag}_{i}(\vec x)v \vec{\s} \vec p
%\vec{\s} \cdot \lp \vec p - \textstyle{\frac{e}{c}} \vec{A}( \vec x, t ) \rp 
\psi_{i}(\vec x)+\! %\frac{e^2 }{2\kappa}
\int\!\!\int\! d^2x d^2x' \frac{:\!\rho(\vec x) \rho(\vec x')\!:}{ 2\kappa |\vec x-\vec x'|} .
\ee
Here $N=4$ is the spin/valley degeneracy, $\psi_i(\vec x)$ and $\psi^{\dag}_i(\vec x)$ describe two-component Dirac fermions on the sublattices $A$ and $B$ of the graphene lattice. The normal-ordered interaction term is written in terms of carrier charge density
$\rho(\vec x)=\sum_{i = 1}^N e\psi^{\dag}_{i}(\vec x)\psi_{i}(\vec x)$ and includes an effective dielectric constant $\kappa$. 

Our main focus here will be on the soft excitations emitted by a photogenerated high-energy e-h pair. To describe this process in the framework of the graphene Hamiltonian, Eq. \eqref{eq:H=H0+V}, we incorporate the optical field in the kinetic-energy term 
%\addCL{via canonical coupling}
%through 
via Peierls substitution 
$\vec p\to \vec p - \textstyle{\frac{e}{c}} \vec{A}( \vec x, t )$, where $\vec{A}( \vec{x}, t )$ describes the optical field. Since optical wavelengths are considerably larger than the photoexcited electron and hole wavelengths, we ignore the position dependence in $\vec{A}( \vec{x}, t )$, treating it as a spatially uniform time-dependent perturbation. 

Before diving into the discussion of soft pairs it is instructive to recall the treatment of photon absorption in the Dirac band \cite{nair2008,stauber2008}. %,mishchenko2008,sheehy2009,juricic2010}.  %mecklenburg2010}. 
The process of
%The rate $W_{0\to1}$ should be compared to the on-shell transition rate describing 
photoexcitation that creates a single hard e-h pair and no soft pairs is pictured in Fig. \ref{fig1}(c). The rate for this process is given by a simple "golden rule" expression, 
\begin{align}\label{eq:on_shell_rate}
W_0 = \frac{2\pi N}{\hbar} \sum_{\vec{k},\vec{k}^\prime} &f_{\vec{k}^\prime}(1-f_{\vec{k}}) \delta_{\vec{k},\vec{k}^\prime}\delta(\epsilon_{\vec{k}}-\epsilon_{\vec{k}^\prime}- h \nu)|M_0|^2,
%\lvert \vec{\sigma} \textstyle{\frac{ev}{c}} \vec{A} \rvert \vec{k}^\prime \rangle\rvert^2,
\end{align}
with $M_0=\langle \vec{k} 
\lvert \vec{\sigma} \textstyle{\frac{ev}{c}} \vec{A} \rvert \vec{k}^\prime \rangle$. Here $f_{\vec{k}}$ and $f_{\vec{k}'}$ is the Fermi-Dirac distribution in the conduction and valence bands 
$\epsilon_{\vec{k}} = \pm v \lvert \vec{k} \rvert$. For brevity, from now on we incorporate the factor $\frac{ev}{c}$ in $\vec{A}$. 
The delta function $\delta_{\vec{k},\vec{k}^\prime}$ arises because the inequality $v\ll c$ allows us to ignore momentum transfer to and from the optical field, as discussed above. We set $f_{\vec{k}^\prime} =1$ and $f_{\vec{k}}=0$, accounting for the fact that the photon energy $h\nu$ is much greater than the smearing of the Dirac point due to temperature and disorder. Assuming linear polarization of the optical field and plugging the two-component Dirac spinor states $|\vec k\ra$,  $|\vec k'\ra$ we find the matrix element
\be\label{eq:matrix_element}
\Big| \langle \vec k' \lvert  \vec{\sigma} \vec{A} \rvert \vec k \rangle \Big|^2=\vec A^2\sin^2(\theta_{\vec k}-\theta_{\vec A}) 
,
\ee 
where the angle $\theta_{\vec k}-\theta_{\vec A}$ describes orientation of the velocities of excited particles relative to the optical field polarization direction.  
We arrive at 
\begin{align}
\label{eq:W_on_shell}
W_0 = \oint\frac{d\theta_{\vec k}}{2\pi} %\frac{e^2 }{c^2} 
\vec{A}^2 \frac{h \nu N}{4 v^2}\sin^2(\theta_{\vec k}-\theta_{\vec A})
%=\frac{e^2 }{c^2} 
=\vec{A}^2 \frac{h \nu N}{8 v^2}
.
\end{align}
The angular dependence $\sin^2(\theta_{\vec k}-\theta_{\vec A})$ indicates that the velocities of excited  pairs are oriented predominantly transversely to $\vec A$, and there are no pairs aligned with $\vec A$. 

\section{Soft pair excitations}

Next we proceed to evaluate the %transition 
rate $W_1$ describing %the process of 
excitation of a primary hard pair and one additional soft pair. We start with general considerations and then %proceed to 
discuss the log divergence. A general expression for this rate, given by the FermiÕs golden rule, reads
\begin{align}\label{eq:W1_Golden_Rule}
&W_1=\frac{2\pi}{\hbar}N^2\sum_{1,1',2,2'}f_{1'}(1-f_1)f_{2'}(1-f_2) |M|^2
%\delta_\epsilon \delta_{\vec k}
\\ \nonumber 
& %\delta_\epsilon=
\times \delta(\epsilon_{\vec k_1}+\epsilon_{\vec k_2}-\epsilon_{\vec k_1'}-\epsilon_{\vec k_2'}-h\nu)
%,\quad 
%\delta_{\vec k}=
\delta(\vec k_1+\vec k_2-\vec k_1'-\vec k_2')
,
\end{align}
where the factors $1-f_1$, $f_{1'}$, $1-f_2$, $f_{2'}$ describe occupancies of the states with momenta $\vec k_1$, $\vec k_1'$, $\vec k_2$, $\vec k_2'$. The transition matrix element $M$ is a sum of two second-order contributions, due to an electron $1$ and a hole $1'$, which differ by the order of the operators describing photon absorption and soft pair creation,
\be\label{eq:matrix_element_M}
M=\tilde V_{\vec q,\omega}\la 1,2|G(\vec p)\vec{\s}\vec A+\vec{\s}\vec A G(\tilde{\vec p})|1',2'\ra
,
\ee
where $G(\vec p)$ and $G(\tilde{\vec p})$ are noninteracting fermion propagators and $\tilde V_{\vec q,\omega}$ denotes the RPA-screened interaction,
\be\label{eq_RPA_V_Pi}
\tilde V_{\vec q,\omega}=\frac{V_{\vec q}}{1-V_{\vec q}\Pi_{\vec q,\omega}}
,\quad
\Pi_{\vec q,\omega}=-\frac{i N}{16\hbar}\frac{\vec q^2}{\sqrt{\omega^2- v^2\vec q^2}}
, %\quad
\ee
with $V_{\vec q}=\frac{2\pi e^2}{\kappa |\vec q|}$ the Coulomb interaction 2D Fourier transform. %From now on we set $v=\hbar=1$, restoring physical units at the end. 
The two terms in Eq. \eqref{eq:matrix_element_M} describe the processes in which photon absorption occurs before and after a pair creation. The virtual states in the two contributions,
Eq. \eqref{eq:matrix_element_M}, are characterized by the off-shell energy values 
$\epsilon_{\vec p}=\epsilon_{\vec k_1}-\omega$, %=\epsilon_{\vec k_1'}+h\nu$, 
$\vec p=\vec k_1-\vec q$ and $\epsilon_{\tilde{\vec p}}=\epsilon_{\vec k_1'}+\omega$, %=\epsilon_{\vec k_1}-h\nu$, 
$\tilde{\vec p}=\vec k_1'+\vec q$
(we use notation from Fig. \ref{fig1}). The Green functions in Eq. \eqref{eq:matrix_element_M} can then be evaluated using the soft pair approximation $\omega\ll \epsilon_{\vec k_1}$, $|\vec q|\ll |\vec k_1|$, giving $G(\vec p)=\frac1{\omega-{\vec v}_1\cdot\vec q}$, $G(\tilde{\vec p})=-\frac1{\omega+{\vec v}_1\cdot\vec q}$ with $\vec v_1\parallel \vec k_1$. 
%\addCL{with $\vec v_1 = \vec{k}_1/|\vec{k}_1|$.} 
% \addLL{(here $\vec n$ is a unit vector pointing along $\vec k_1$).} 

To simplify the expression in Eq. \eqref{eq:W1_Golden_Rule} we transform the sum over momenta by splitting the delta functions as
\begin{align}\nonumber
\delta(&\epsilon_{\vec k_1}+\epsilon_{\vec k_2}-\epsilon_{\vec k_1'}-\epsilon_{\vec k_2'}-h\nu)\nonumber\\
&=\int d\omega \delta(\epsilon_{\vec k_2}-\epsilon_{\vec k_2'}-\omega)\delta(\epsilon_{\vec k_1}-\epsilon_{\vec k_1'}+\omega-h\nu),\nonumber\\
\delta(&\vec k_1+\vec k_2-\vec k_1'-\vec k_2')=\nonumber\\ %\nonumber
&=\int d^2q\delta(\vec k_2-\vec k_2'-\vec q)\delta(\vec k_1-\vec k_1'+\vec q)
,
\end{align}
where $\omega$ and $\vec q$ are energy and momentum of soft pairs.  The sum over $\vec k_2$ and $\vec k_2'$ can then be expressed through polarization function using the familiar  identity \cite{hwang2007,wunsch2006} 
\be
{\rm Im}\,\Pi_{\vec q,\omega}=-N\pi \sum_{\vec k_2} F_{\vec k_2,\vec k_2'}(f_{\vec k_2'}-f_{\vec k_2})\delta(\epsilon_{\vec k_2}-\epsilon_{\vec k_2'}-\omega),
\ee 
where $F_{\vec k_2,\vec k_2'}$ are coherence factors that are implicit in Eq. \eqref{eq:matrix_element_M}. Plugging it into Eq. \eqref{eq:W1_Golden_Rule} we obtain
\be
\label{eq:pen_ult_step}
W_1 = -\frac{8 N}{\hbar} \sum_{\vec{k}_1, \vec{q}} ( N_\omega +1)
|\tilde{V}_{\vec{q}, \omega}|^2
\Im\Pi_{\vec{q}, \omega} \left\lvert\frac{v q_{\parallel} \la 1 | \vec{\sigma}\vec{A} | 1' \ra}{\omega^2-v^2 q_{\parallel}^2}\right\rvert^2
,
\ee
where $\vec k_1$ is the momentum at which the primary hard pair is excited, $\vec q$ is the momentum transferred to the soft pair and $\omega = h \nu - 2 v \lvert \vec{k}_1\rvert - v q_{\parallel}$ is the energy of the soft pair, where $q_{\parallel}$ denotes the component of vector $\vec{q}$ parallel to $\vec{k}_1$. The latter relation follows from the energy conservation $\omega=h\nu-\epsilon_{\vec k_1}-\epsilon_{\vec k_1+\vec q}$, see Fig. \ref{fig1}. %LL \addLL{[OK?]}

Tackling the double-log divergence in the rate $W_1$, Eq. \eqref{eq:pen_ult_step}, is facilitated by expressing the integral over $\vec k_1$ through an integral over the soft pair frequency $\omega=h\nu-2v|\vec k_1|-v q_\parallel$. 
%LL \addLL{[simplify]} 
This is done by going to polar coordinates as %writing the sum over $\vec{k}_1$ as
\be
\label{eq:k1}
\sum_{\vec{k}_1}\dots  = \int\limits_{0}^{\infty}\int\limits_{0}^{2\pi} \frac{k_1 d k_1 d \theta_1}{(2\pi)^2} \dots %\frac{d \theta_{\vec k_1}}{2\pi} \dots 
\approx \frac{h \nu}{8\pi v^2} \int\limits_{-\infty}^{\infty} 
d\omega \int\limits_{0}^{2\pi}\frac{d \theta_1}{2\pi} \dots, %\dots
\ee
where we used the soft pair approximation $\omega\ll h\nu$ to introduce a constant density of states at half the photon energy $\epsilon=h\nu/2$. We note that, while the integral in Eq. \eqref{eq:k1} runs over $-\infty<\omega<\infty$, the physical values $\omega$ are smaller than half the photon energy $h\nu/2$. We will therefore use $h\nu/2$ as a UV cutoff in integration over $\omega$ whenever necessary. 

Inserting Eq. \eqref{eq:k1} into Eq. \eqref{eq:pen_ult_step} we see that, in complete analogy with the on-shell rate \eqref{eq:W_on_shell}, the dependence on $\vec k_1$ orientation relative to $\vec A$, namely the angle $\theta_1-\theta_{\vec A}$, originates only from the matrix element \eqref{eq:matrix_element}. It will shortly be clear that the soft pairs are emitted in the direction which is collinear with the hard pair direction, forming two ``jets'' directed along $\vec k_1$ and $-\vec k_1$.
%The sine function in Eq. \eqref{eq:matrix_element} therefore indicates that the jets are oriented  predominantly transversely to $\vec A$. 
Angular integration 
$\int_{0}^{2\pi}\frac{d \theta_1}{2\pi} \Big| \langle 1 \lvert  \vec{\sigma} \vec{A} \rvert 1' \rangle \Big|^2 = \frac12 \vec{A}^2$ 
is therefore equivalent to averaging over all possible orientations of the jets. Angular distribution %$\sin^2(\theta_1-\theta_{\vec A})$ 
of the jets thus replicates that of the primary pairs, as expected. 
%LL [{\bf \addLL{(add discussion of this point to the main text?)}}]

It will be convenient to factor out the hard-pair rate $W_0$ [see Eq. \eqref{eq:W_on_shell}] and write the transition rate in Eq. \eqref{eq:pen_ult_step} as an integral over the frequency and momentum of soft pairs as
\be\label{eq:W01_w_q}
W_1 = -\frac{4}{\pi\hbar} W_0\!\!  \int\limits_{-\infty}^\infty\!\!  d\omega \sum_{\vec{q}} (N_\omega +1) 
|\tilde{V}_{\vec{q}, \omega}|^2
%\Im\Pi(\vec{q}, \omega) %\left\lvert
\frac{\Im\Pi_{\vec{q}, \omega} v^2 q_{\parallel}^2 
}{(\omega^2-v^2 q_{\parallel}^2)^2}, %\right\rvert^2
\ee
where %we factored out  the hard-pair rate 
%$W_0=\frac{e^2\vec A^2 h\nu}{c^2}\frac{N}8$ (see Eq. \eqref{eq:W_on_shell}) and
we  introduced $q_\parallel$, the component of $\vec q$ parallel to the hard pair velocity direction. 
The Bose function $N_\omega =\frac1{e^{\beta\omega}-1}$ 
describes thermal broadening of the Dirac point.
%LL \addLL{[fix fences in line 198]}. 
In the limit $\omega\gg k_BT$ the dependence $N_\omega +1$ can be approximated as the Heaviside function $\Theta(\omega>0)=1$, $\Theta(\omega<0)=0$. In this case the soft pairs have positive energies. %, as discussed in the main text [see paragraph following Eq. (11)].  
%LL \addLL{[fix it in galley proofs]} 

Interestingly, while the dependence $N_\omega+1$ at $T>0$ does give rise to thermal smearing of the Dirac point, it does not in itself regularize the infrared divergence in Eq. \eqref{eq:W01_w_q}. The origin of this peculiar behavior can be seen from the identity $N_\omega +N_{-\omega}+1=0$, which allows us to reduce integration over $-\infty<\omega<\infty$ to integration over $0<\omega<\infty$ through
\begin{align}
\label{eq:bose_identity}
%\int_{-\infty}^\infty (N_\omega +1) F(\omega) d\omega
%= \int_0^{\infty} F(\omega) d\omega
&\int_{-\infty}^\infty (N_\omega +1) F(\omega) d\omega \nonumber\\
&=\int_{-\infty}^{0} (N_\omega+1)F(\omega) d\omega +\int_0^{\infty} (N_\omega+1) F(\omega) d\omega\nonumber\\
&=\int_0^{\infty}(N_{-\omega}+N_{\omega}+1) F(\omega) d\omega+\int_0^{\infty} F(\omega) d\omega\nonumber\\
 &= \int_0^{\infty} F(\omega) d\omega,
\end{align}
valid for any even integrable function $F(\omega)$. Temperature dependence due to $N_\omega+1$ therefore drops out for any contribution expressed through an even function of $\omega$, such as Eq. \eqref{eq:W01_w_q}. We note parenthetically that, while explicit temperature dependence drops out from the Eq. \eqref{eq:pen_ult_step} due to the above property of the function $N_\omega$, finite-$T$ effects will enter implicitly through the polarization function $\Pi_{\vec{q},\omega}$ and the Fermi-Dirac distribution factors in Eqs. \eqref{eq:on_shell_rate} and \eqref{eq:W1_Golden_Rule}. These effect will enable interband and intraband on-shell processes, thereby increasing the observed total transition rate.

\section{The double-log divergence in the rate $W_1$}

To analyze the double-log divergence in $W_1$ transition rate it is convenient to nondimensionalize this quantity by introducing the mean number of excited pairs, 
%defined by a relation 
%Ensemble-averaged number of secondary pairs excited by a single photogenerated hard e-h pair can be written as a ratio of the rates $W_{0\to1}$ and $W_{\text{on-shell}}$:
\be\label{eqN_p}
\la N_{\rm p}\ra=\frac{W_1}{W_0}
.
\ee
Such a relation between a one-pair excitation rate and the mean number of pairs is valid provided that different pairs are statistically independent and thus obey Poisson statistics. 
The rates $W_1$ and $W_0$ are related through Eq. \eqref{eq:W01_w_q}, giving a simple expression for $\la N_{\rm p}\ra$. 

To investigate the dependence in Eq. \eqref{eq:W01_w_q} we define a dimensionless coupling constant,
\be
\label{eq:eta_and_g}
g = \frac{\pi N \alpha}{8} %(1-e^{-2H|\vec q|})
= \frac{\pi N e^2}{8\kappa \hbar v},  %(1-e^{-2H|\vec q|})
\ee
where $\alpha=\frac{e^2}{\hbar v\kappa}$ is the conventional ``fine-structure'' parameter, and the number of spin/valley flavors $N$ is incorporated in $g$ for later convenience. 

In anticipation of an IR divergence, it is useful to introduce a physically motivated IR regularization in the expressions for $\tilde V_{\vec q,\omega}$ and $\Pi_{\vec q,\omega}$. 
To that end we consider a narrow-gapped Dirac fermion dispersion $\epsilon_{\vec k}=\pm \sqrt{v^2\vec k^2+\Delta^2}$ and describe the soft pair response with the help of a polarization function,
\be
\label{eq:physical_polarization}
\Pi_{\vec{q},\omega}=-\frac{iN\vec q^2}{16\hbar }\frac1{\sqrt{\omega^2-v^2\vec q^2-\Delta^2}},
\ee
which we will use here instead of the one in Eq. \eqref{eq_RPA_V_Pi} that describes gapless graphene. The main effect of such regularization is to suppress pair production with energies below $\Delta$ and small wave numbers $q<k_0=\Delta/v$. % and energies below $\Delta= v k_0$. 

We note parenthetically that the gap parameter $\Delta$, which serves as a vehicle to regulate the IR behavior of perturbation theory, may account for different effects in the system. In particular, it may describe the actual gap in the bandstructure or serve as a proxy for the detector energy resolution. Finite $\Delta$ also introduces a deviation from a linear dispersion relation, providing a regularization for the logs arising at high energies due to angular integration (see below). Plugging Eq. \eqref{eq:physical_polarization} %this expressions 
for $\Pi_{\vec q,\omega}$ into $\tilde V_{\vec q,\omega}=\frac{V_{\vec q}}{1-\Pi_{\vec q,\omega}V_{\vec q}}$ and using the identity \eqref{eq:bose_identity}, 
we write
\begin{align}
\nonumber
&\la N_{\rm p}\ra=
\int_0^\infty  d\omega \int\frac{d^2q}{(2\pi)^2} %\sum_{\vec{q}} %(N_\omega +1) 
\,\frac{2^6 g^2}{\pi N v}
\frac{\sqrt{\frac{\omega^2}{v^2}-\vec q^2-k_0^2}}{\frac{\omega^2}{v^2}-\vec q^2-k_0^2+g^2\vec q^2} 
\\ \label{eq:app_transition_rate_num}
&\times\Theta\lp \frac{\omega^2}{v^2}-\vec{q}^2-k_0^2 \rp
\frac{q_{\parallel}^2}{(\frac{\omega^2}{v^2}- q_{\parallel}^2)^2}
,
\end{align}
where the Heaviside step function %$\Theta(\omega^2-v^2|\vec{q}|^2)$ 
comes from %the 
$\Im\Pi_{\vec{q}, \omega}$. % term. 
The quantity under the integral scales as $q^{-3}$, leading to an IR divergence at $k_0\to0$. 
Importantly, since the regularization $k_0$ affects only the soft pair response function $\Pi_{\vec q,\omega}$, it does not enter the last term originating from the carrier dynamics at high energies which are far above the gap $\Delta$ in the Dirac band. It is this term 
that will generate an additional log divergence. 
%LL \addLL{[OK?]}

The expression in Eq. \eqref{eq:app_transition_rate_num} can be simplified by writing  $q_{\parallel}=|\vec q|\cos\theta$ and integrating over $\theta$ using the identity
\be
\oint \frac{d\theta \cos^2\theta}{(a-\cos^2\theta)^2}=-\p_a\oint \frac{d\theta \cos^2\theta}{a-\cos^2\theta}=\frac{\pi}{a^{1/2}(a-1)^{3/2}}
, 
\ee
with $a=\frac{\omega}{vq}$. 
The integral is dominated by the angles in the regions of size $\delta\theta\approx \sqrt{\frac{\omega^2}{v^2q^2}-1}$ centered at $\theta=0$ and $\pi$, which become very narrow in the regime of interest $\frac{\omega}{vq}\to 1$. The small values of $\delta\theta$ indicate that soft pairs have a sharp angular distribution peaked along the hard pair direction. 
Setting the limits of integration over $\omega$ and $q$ in agreement with the Heaviside function we write
\begin{align}
\nonumber %app_transition_rate_num}
\la N_{\rm p}\ra=
& \int_{\Delta}^\infty \! d\omega \int_0^{q_\omega}  \! qdq   %\sum_{\vec{q}} %(N_\omega +1) 
\,\frac{16 g^2}{\pi^2 N}
\frac{%\Theta(\omega^2-v^2|\vec{q}|^2) 
\sqrt{\frac{\omega^2}{v^2}- q^2-k_0^2}}{\frac{\omega^2}{v^2}- q^2-k_0^2+g^2 q^2}
\\ \label{eq:N_w_q} 
& \times \frac{ q^2}{\omega \lp\frac{\omega^2}{v^2}- q^2\rp^{3/2}}
%\frac{q_{\parallel}^2}{(\frac{\omega^2}{v^2}- q_{\parallel}^2)^2}
,
\end{align}
with $q_\omega=\frac1{v}\sqrt{\omega^2-\Delta^2}$.
To understand the properties of the integral over $q$ in Eq. \eqref{eq:N_w_q} it is convenient to temporarily suppress regularization and set $k_0=0$. This gives an expression that diverges logarithmically at the upper limit $q=\omega/v$ because of the first term in the numerator and the last term in the denominator.  The log divergence is cut off after reinstating $k_0\ne 0$, giving
\be
\label{eq:N_w}
\la N_{\rm p}\ra=
\int_{\Delta}^\infty d\omega 
\, \frac{16}{\pi^2 N }\frac{\ln\frac{\omega}{\Delta}}{\omega}
,
\ee
where we neglected nonlogarithmic contributions assuming $\ln\frac{\omega}{\Delta}\gg 1$. The expression under the integral gives the spectrum of emitted pairs, Eq. \eqref{eq:spectrum}. Integration over $\omega$ then generates the double-log dependence of Eq. \eqref{eq:Nave}. Interestingly, the number of secondary pairs produced per single absorbed photon, given in Eq. \eqref{eq:Nave}, as well as their spectrum in Eqs. \eqref{eq:Nave} and \eqref{eq:N_w}, depends on the number of flavors $N$ but not on the interaction strength $\alpha$.

As noted above the log factor in Eq. \eqref{eq:N_w} arises after integration over $q$ near the singularity at $q=\omega/v$ in the denominator of Eq. \eqref{eq:N_w_q}. This singularity, in turn, originates from integrating $1/(\omega^2-v^2q_\parallel^2)^2$ over the angle between the soft and hard pair momenta. This  confirms that the second log in the double-log dependence is a signature due to the near collinear character of soft pairs. 

The angular log divergence will of course be sensitive to any deviation of the dispersion from a linear dependence, which will generate an IR cutoff in this divergence. While in the above analysis the dominant nonlinear effects in dispersion arise due to the gap $\Delta>0$, in practice there are other effects that can distort Dirac cones. One such effect is the trigonal warping interaction. Another potentially relevant effect is the energy dependence of the graphene carrier velocity arising due to electron interactions reshaping Dirac cones into funnels (see Ref. \cite{elias2011}). Although this effect is log-divergent we ignored it in our analysis of the pair production rate since it is subleading to the double-log divergence effects. It may, however, play a role in regulating the log divergence in angular integration.

\section{The effect of screening }

Next we consider the effect of screening by a gate. The latter can be incorporated by replacing the $1/r$ interaction in Eq. \eqref{eq:H=H0+V} with $V(\vec x-\vec x')=\frac{e^2}{\kappa |\vec x-\vec x'|}-\frac{e^2}{\kappa |\vec x-\tilde{\vec x}'|}$, 
%-\frac{e^2}{\kappa \sqrt{(\vec x-\vec x')^2+4h^2}}$, 
where the second term accounts for the image charges due to the gate. Below we will need the Fourier transform of this interaction (here $H$ is the distance to the gate):
\be
V_{\vec q}=\int d^2x e^{-i\vec q(\vec x-\vec x')} V(\vec x-\vec x') %\frac{e^2}{\kappa}\lp \frac1{ |\vec x-\vec x'|}-\frac1{ |\vec x-\tilde{\vec x}'|}\rp
=\frac{2\pi e^2}{\kappa|\vec q|}\lp 1-e^{-2H|\vec q|}\rp.
\ee 
%where $H$ is the distance to the gate. 
We will see that the main effect of the gate 
%on the soft pair generation 
is to suppress pair production with wave numbers %momenta and frequencies 
smaller than $k_H=1/(2H)$. % and $\hbar v/2h$, respectively, 
One might expect that this suppression translates into an effective low-energy cutoff value $\Delta_H= v/(2H)$ which, if greater than $\Delta$, will replace it in Eq. \eqref{eq:spectrum}. As we will see, this naive expectation is incorrect and the actual situation is %considerably 
more interesting. Namely, only one of the two logs in the double-log dependence of Eq. \eqref{eq:Nave} will be cut at $\Delta_H$ whereas the other log will be cut at a much smaller energy scale. Such a behavior arises because screening, while suppressing the contribution of pairs with small wave numbers, $q<k_H$, has absolutely no impact on the singularity in the angular integration over $\theta$ which receives contributions from pairs with all $q$ values, large and small. As a result the corresponding log divergence is insensitive to screening. 
%LL \addLL{check}

%due to a peculiar interplay of the scales $\Delta_H$ and $\Delta$ in the soft pair excitation dynamics. \addLL{[Reason: angular divergence, ref to Eq.}

From a technical standpoint, analysis of the screened interaction requires only minor modifications of the above discussion. The number of excited pairs in this case is still given by Eq. \eqref{eq:N_w_q}, however the dimensionless coupling $g$ is now replaced with %Initial analysis follows the same steps as 
\be
g_q= \frac{\pi N e^2}{8\kappa \hbar v}  (1-e^{-v|\vec q|/\Delta_H})
.
\ee
It is instructive to consider the change in the number of emitted pairs $\delta\la N_{\rm p}\ra$ due to introduction of  screening. 
%Subtracting the integrals and 
Evaluating the change of the expression under the integral in Eq. \eqref{eq:N_w_q} 
%due to introduction of screening 
with the help of the identity
\begin{align}
&\frac{g_q^2 q^2}{\frac{\omega^2}{v^2}- q^2-k_0^2+g_q^2 q^2}
-\frac{g^2 q^2}{\frac{\omega^2}{v^2}- q^2-k_0^2+g^2 q^2}
\\ \nonumber
&
=\frac{(g_q^2-g^2) q^2\lp\frac{\omega^2}{v^2}- q^2-k_0^2\rp}{\lp\frac{\omega^2}{v^2}- q^2-k_0^2+g_q^2 q^2\rp \lp\frac{\omega^2}{v^2}- q^2-k_0^2+g^2 q^2\rp},
\end{align}
we note that, since the last term in the numerator vanishes at the upper limit of integration  $q_\omega=\frac1{v}\sqrt{\omega^2-\Delta^2}$, the integral over $q$ in the expression for $\delta\la N_{\rm p}\ra$  no longer diverges at the upper limit when $k_0$ tends to zero. It is therefore safe to set $k_0=0$ in the inner integral and analyze the expression
\begin{align}
\nonumber %app_transition_rate_num}
&
\delta \la N_{\rm p}\ra \!=\!
\int\limits_{\Delta}^\infty \! \frac{d\omega}{\omega} \! \int\limits_0^{\frac{\omega}{v}} \! %qdq   %\sum_{\vec{q}} %(N_\omega +1) 
\,\frac{16 }{\pi^2 N }
\frac{(g_q^2-g^2) q^3 dq 
}{\lp \frac{\omega^2}{v^2}- q^2+g_q^2 q^2\rp \lp\frac{\omega^2}{v^2}- q^2+g^2 q^2\rp}
\\ \label{eq:delta_N_w_q} 
&
=\!\int\limits_{\Delta}^\infty \!\frac{d\omega}{\omega} \int\limits_1^\infty \!  \frac{16}{\pi^2 N x }
\frac{(g^2_\omega(x)-g^2)dx}{ \lp x^2-1+g^2\rp \lp x^2-1+g^2_\omega(x)\rp}
,
\end{align}
where we made a substitution $q(x)=\frac{\omega}{vx}$, $1<x<\infty$, and defined $g_\omega(x)=g(1-e^{-\omega/x\Delta_H})$. In this expression 
the numerator is exponentially small when $x\ll \omega/\Delta_H$. In the opposite limit, $x\gg \omega/\Delta_H$, we have $g_\omega(x)\approx g\omega/(x\Delta_H)\ll g$. Integral over $x$ is dominated by values $0<x-1\ll 1$, giving 
\be
-\frac{16 }{\pi^2 N \omega }\ln\frac{\Delta_H}{\omega}
\ee
provided $\omega$ is smaller than $\Delta_H$. Since at larger $\omega$ % values 
the integral over $x$ drops rapidly, we can estimate $\delta \la N_{\rm p}\ra$ as
\be
\label{eq:Np_estimate}
\delta \la N_{\rm p}\ra \!=\! -\int_{\Delta}^{\Delta_H}\!\!\frac{16 }{\pi^2 N  }\ln\frac{\Delta_H}{\omega}\frac{d\omega}{\omega}=-\frac{8 }{\pi^2 N} \ln^2 \frac{\Delta_H}{\Delta}
,
\ee
a result valid for not too large gate-graphene separation, $2Hk_0\ll 1$. In the opposite limit of a large distance to the gate, $2Hk_0\gg 1$, the effects of interaction screening by the gate are inessential $\delta\langle N_p\rangle \ll \langle N_p\rangle$. Focusing on the case of a proximal gate and using the result in Eq. \eqref{eq:Np_estimate} we find
\begin{align}\nonumber
\la N_{\rm p}\ra = \frac{8 }{\pi^2 N}\lp \ln^2\frac{vk_{\nu}}{\Delta} - \ln^2\frac{\Delta_H}{\Delta} \rp 
.
\end{align}
After rearranging the logs this quantity can be written as a QED-like double-log dependence with two cutoffs: 
\be\label{eq:Np=loglog}
\la N_{\rm p}\ra =\frac{8 }{\pi^2 N} \ln\frac{vk_{\nu}}{\Delta_H}\ln\frac{vk_{\nu}}{\mu}
,
\quad
\mu = \frac{\Delta^2}{\Delta_H}
.
\ee
Interestingly, while screening by a gate reduces the value $\la N_{\rm p}\ra$, as expected, it does not eliminate the dependence on the Dirac gap $\Delta$ even when this gap is much smaller than $\Delta_H$. The dependence on $\Delta$ survives in one of the two logs, playing a role similar to the photon mass cutoff in the QED double-log dependence in Eq. \eqref{eq:sudakov}. % which persists even for $Hk_0\ll 1$.

One remarkable property of the result for $\la N_{\rm p}\ra$, Eq. \eqref{eq:Np=loglog}, is its universality: the prefactor $8/\pi^2 N$ depends only on the number of flavors but not on  %coupling 
the fine-structure %constant 
parameter value $\alpha$. The dependence on $\alpha$ may of course appear in the cutoffs of the two logs, yet it completely drops out from the prefactor. We also stress that we have not used any form of $1/N$ expansion in the derivation, and thus the $1/N$ dependence in Eq. \eqref{eq:Np=loglog} is valid for all $N$, large or not too large. 

Another peculiar property of the $\la N_{\rm p}\ra$ double-log dependence is that it is surprisingly insensitive to the effects that readily provide IR regularization in other cases. For instance, the divergence in the gapless limit $\Delta\to0$ survives at a finite temperature or under screened carrier-carrier interaction. It remains to be seen whether this divergence can be cut off by subleading effects, although we believe this is unlikely to be the case.

\section{Conclusions}

Summing up, fast carrier dynamics %in graphene 
can lead to emission of multiple soft e-h pairs with a characteristic power-law energy distribution. Our analysis reveals an analogy between this effect and %soft photon emission in 
Bremsstrahlung radiation due to fast charge dynamics. %, a signature QED effect. 
The main difference between this signature QED effect and our e-h pair production processes stems in the coupling strength. While for Bremsstrahlung radiation the coupling strength is %$1/137$, i.e. 
quite weak, it is an order-one effect for e-h pair production in graphene. We therefore expect the soft pairs produced under photoabsorption to gain prominence in the photoexcitation cascade. 
%Theory of this effect is analogous to  This behavior 
A discussion of the detection of secondary soft pairs in photoabsorption, in particular their energy spectrum and angular distribution, can be found in Ref. \cite{lewandowski2018}. Multiple pair emission is of interest in relation to searching for materials exhibiting strong hot-electron effects and/or carrier multiplication under photoabsorption. While our analysis focuses on graphene, the conclusions apply more broadly to other Dirac materials with linear or nearly linear band dispersion, either 2D or 3D. 

\begin{acknowledgments}
We acknowledge support of the Center for Integrated Quantum Materials (CIQM) under NSF Award No. DMR-1231319, and MIT Center for Excitonics, an Energy Frontier Research Center (EFRC) funded by the U.S. Department of Energy,
Office of Science, Basic Energy Sciences under Award No. DE-SC0001088.
\end{acknowledgments}

\end{document}